# Observation of an Effective Magnetic field for Light

Lawrence D. Tzuang,[1] Kejie Fang,[2] Paulo Nussenzveig,[1,3] Shanhui Fan,[2] and Michal Lipson[1,4,*]

[1]*School of Electrical and Computer Engineering, Cornell University, Ithaca, New York 14853, USA*

[2]*Department of Electrical Engineering, Stanford University, Stanford, California 94305, USA*

[3]*Instituto de Física, Universidade de São Paulo, P.O. Box 66318, 05315-970 São Paulo, Brazil*

[4]*Kavli Institute at Cornell for Nanoscale Science, Cornell University, Ithaca, New York 14853, USA*

*Corresponding author: ml292@cornell.edu*

**Photons are neutral particles that do not interact directly with a magnetic field. However, recent theoretical work[1,2] has shown that an effective magnetic field for photons can exist if the phase of light would change with its propagating direction. This direction-dependent phase indicates the presence of an effective magnetic field as shown for electrons experimentally in the Aharonov-Bohm experiment. Here we replicate this experiment using photons. In order to create this effective magnetic field, we construct an on-chip silicon-based Ramsey-type interferometer[3-7]. This interferometer has been traditionally used to probe the phase of atomic states, and here we apply it to probe the phase of photonic states. We experimentally observe a phase change, i.e. an effective magnetic field flux from 0 to $2\pi$. In an Aharonov-Bohm configuration for electrons, considering the device geometry, this flux corresponds to an effective magnetic field of 0.2 Gauss.**

The interaction of light and magnetic field would enable critical non-reciprocal devices such as isolators. Since photons are neutral particles, their interaction with magnetic field relies on using magneto-optical materials. Recently, there were demonstrations of on-chip isolators[8-10] and topologically protected edge modes[11-15] based on magneto-optical materials. However, these materials are difficult to integrate on-chip and the magneto-optic effect is weak in near-IR and visible domain. This leads to the fundamental question of whether one can generate an effective magnetic field directly coupled with photons in the optical domain while not limited to magneto-optical materials.

As shown by Fang et al.[1,2], an effective magnetic field for photons could be created, if one could break the reciprocity of light such that its phase would depend on its propagation direction. The link between the magnetic field ($\vec{B}$) (and its associated gauge potential ($\vec{A}$)) to the direction-dependent phase is equivalent to the Aharonov-Bohm (AB) effect[16] for electrons where the electrons acquire a direction-dependent phase ($\phi = \frac{e}{h}\int_{r}^{r'} \vec{A} \cdot d\hat{r}$, where $B = \nabla \times \vec{A}$, e is the unit charge, and h is the Planck constant) in the presence of a magnetic field. Recently, the effective magnetic field for radio-frequency photons (RF) was observed using a photonic Aharonov-Bohm interferometer[17]. The demonstration of such an effect in the on-chip optical domain provides a new functionality for on-chip light manipulation.

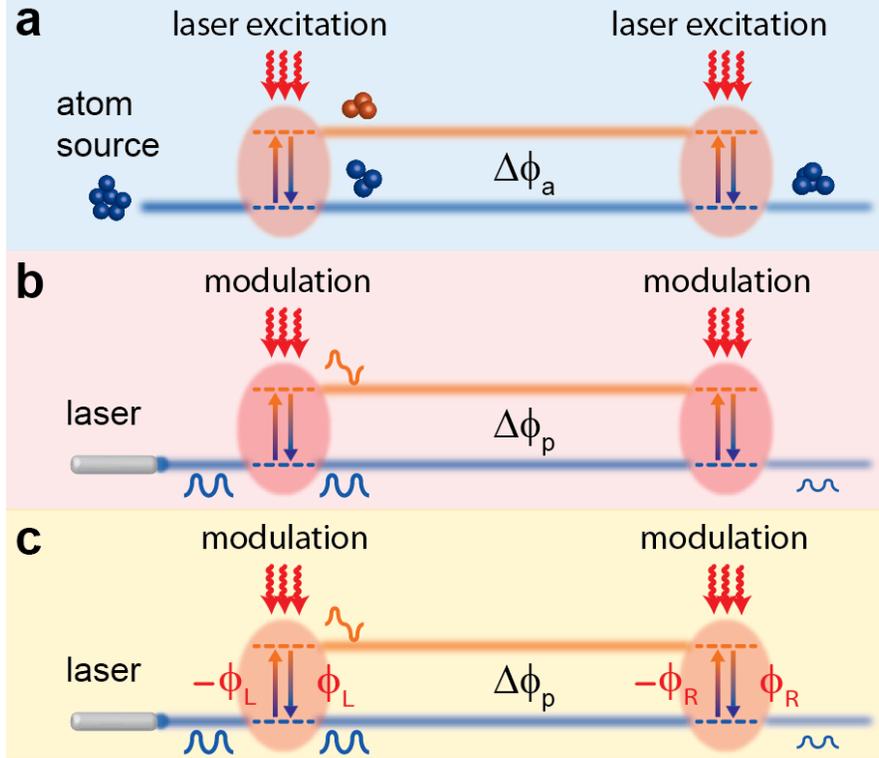

Fig. 1 **Effective magnetic field for light using a Ramsey-type interferometer.** Illustration of (a) an atomic Ramsey interferometer, (b) a photonic Ramsey Interferometer, and (c) a photonic Ramsey interferometer where the two modulators have different phases $\phi_L$ and $\phi_R$.

Here we probe the phase of light using a Ramsey-type interferometer[3-7]. The basic form of a Ramsey-type interferometer is shown in Fig. 1a. In an atomic Ramsey interferometer, as an atom in the ground state enters the interferometer, the first laser (left) interacts with it, and the atomic state is rendered in a linear superposition of the ground and excited state. These two states have different propagation phases ($\Delta\phi_a$, due to rotation and gravitation for example). A second laser excitation, in phase with the first, once more transforms the atomic ground and excited states into linear superpositions. Thus, the probability of finding an atom exiting the interferometer in the ground state exhibits an interference profile depending on $\cos(\Delta\phi_a)$. In a photonic Ramsey interferometer (Fig. 1b), we replace the two atomic states and laser excitations with two photonic states (in our case even- and odd-modes in a waveguide (see below)) and modulators, respectively. As light in the even-mode (ground state) enters the interferometer, the first modulator (left) induces a refractive index perturbation and couples a portion of light in the even-mode to the odd-mode (excited state). Note that this transition is classical in contrast to the atomic Ramsey interferometer. Following the excitation (i.e., coupling), similar to the atomic case, the propagating light is in a superposition of the even and odd-mode. The two modes experience different phases ($\Delta\phi_p$) owing to their different propagation constants. A second modulator (right) couples light in the odd-mode back into the even-mode and light exiting the interferometer exhibits an interference profile, as in the atomic version but now depending on $\cos(\Delta\phi_p)$.

We use the Ramsey interferometer to probe the phase and break the reciprocity of light thus inducing an effective magnetic field. This is achieved if the two modulators have different phases $\phi_L$ and $\phi_R$ (Fig. 1c). When inducing couplings, modulators impart their phases on photons. With respect to the phase of the local oscillator that drives the modulator, the imparted phase on photons is negative (positive) if excitation (de-excitation) occurs[1]. If the phases of both modulators are identical (Fig. 1b), then the total imparted phases are cancelled. However, if the modulators have different phases (Fig. 1c), these imparted phases are detected and the transmission becomes direction-dependent. When light enters the interferometer from the left (right), the output of the interferometer is proportional to $\cos(\Delta\phi_p-\phi_L+\phi_R)$ ($\cos(\Delta\phi_p-\phi_R+\phi_L)$). This non-reciprocal transmission is a result of an effective magnetic flux, where $B_{flux}=\phi_L-\phi_R$ (see Ref. 1).

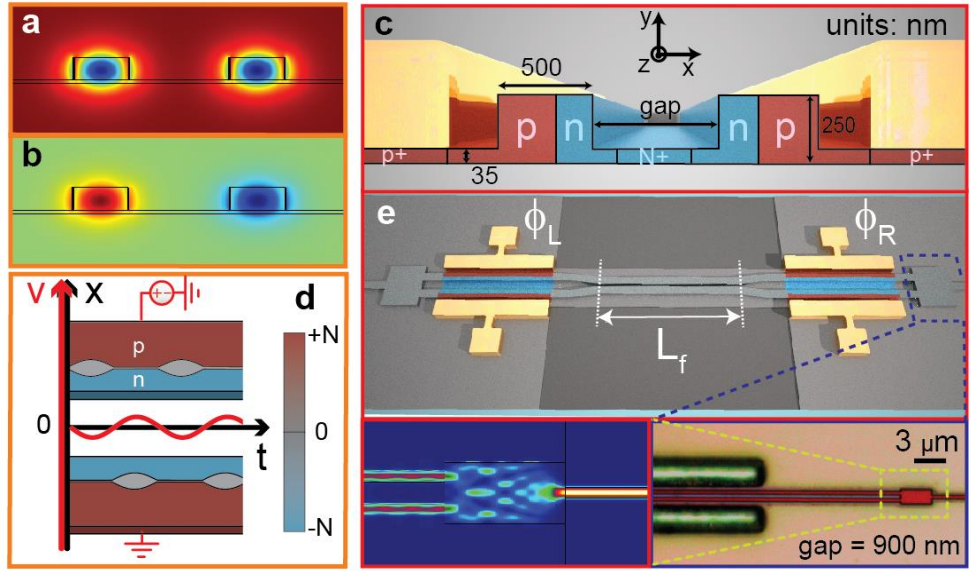

Fig. 2 **Ramsey-type interferometer design and fabrication.** Simulated mode profile for both (a) the even-mode and (b) the odd-mode which coexist in a silicon coupled-waveguides structure. (c) Cross-sectional view of the coupled-waveguides. A set of p-n and n-p diodes is doped in this coupled-waveguides to modulate the refractive index. (d) Top view of the carrier density (N) distribution of the coupled-waveguide along the x-axis (the slab is omitted). The width of the depletion region (gray) changes over time as a sinusoidal signal is applied to the diodes. The applied sinusoid voltage (V) is shown in red. (e) Illustration of a photonic Ramsey interferometer implemented by a silicon coupled-waveguides structure. The insets show the microscope image and the simulated light transmission of a pair of multi-mode interference devices located at the outer ends of the interferometer.

We implement the photonic Ramsey interferometer by using the supermodes (even- and odd-modes) of a silicon coupled-waveguides structure. The mode profiles are shown in Figs. 2a and 2b, and the dimensions of the structure are shown in Fig. 2c. The modulators are formed by embedding pn and np diodes in the waveguides (Fig. 2c). Figure 2d shows the top view of the carrier distribution under an applied sinusoidal voltage (red). The width of the depletion region (gray) changes as signal is applied, which induces a change in the refractive index of the coupled-waveguides[18, 19]. The pn-np configuration[20] ensures that at any time instance, only one side of the coupled-waveguides experiences a depletion width change, which enables coupling between the supermodes. Fig. 2e illustrates the overview of the interferometer. The two modulators are identical, and only their modulation phases are different ($\phi_R$ and $\phi_L$). The length of both modulators is 3.9 mm, which in simulation provides an equal probability (50 %) to populate both the two supermodes. The gap of the coupled-waveguides varies along the interferometer. At the edges where the modulators are placed, it is equal to 900 nm to separate the two supermodes in frequency by a few GHz in the optical c-band. In the center, the gap is tapered (the taper length is 100 μm) down to 550 nm and extends for a distance $L_f$ such that the two supermodes experience different effective indices $\Delta n_{eff}$, and the phase difference between the two supermodes becomes $\Delta k \times L_f$ ($\Delta k = 2\pi \Delta n_{eff}/\lambda$, and $\lambda$ is the optical wavelength). Here, $L_f$ varies from 175 to 350 μm for different fabricated devices. We also place multi-mode interference (MMI) devices at each ends of the interferometer so that only even-mode enters and exits the interferometer. A microscope image and a simulated power distribution of the MMI are shown in the insets of Fig. 2e.

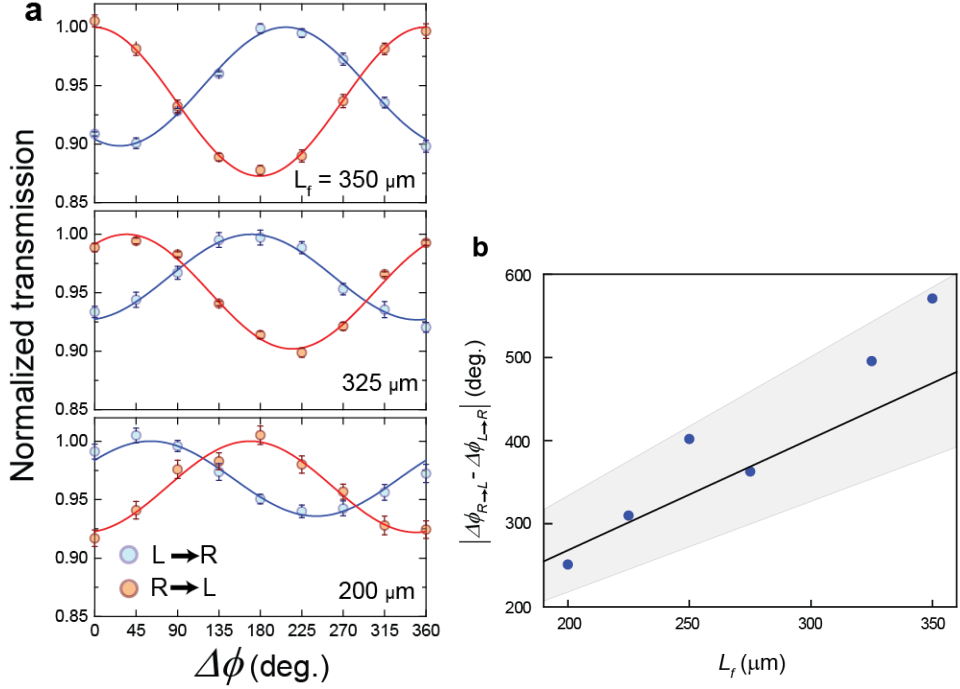

Fig. 3 **Effective magnetic field experiment.** (a) Examples of the measured (circles) and theoretically fitted (solid lines) optical transmission for light traveling from left to right (L→R, blue) and right to left (R→L, red) for devices with different $L_f$ as a function of the phase difference between the two signals applied to the modulators ($\Delta\phi = \phi_L - \phi_R$). The wavelength used is 1570 nm. (b) Measured (circles) and theoretical (Solid) difference between the $\Delta\phi$ when the transmission is maximum for L→R ($\Delta\phi_{L \to R}$) and R→L ($\Delta\phi_{R \to L}$) versus different $L_f$ values. The gray region indicates the error of the theory curve when a 5 % process variation is introduced.

We experimentally observe non-reciprocal fringe patterns indicating the existence of an effective magnetic flux from 0 to $2\pi$. Fig. 3a show the optical transmission of our devices when light is propagating from left to right (L→R) and right to left (R→L). Two synchronized sinusoidal RF signals are applied such that $\phi_L$ and $\phi_R$ are correlated. We choose $\lambda = 1570$ nm to match the modulation frequency ($f_M = 4$ GHz) to the frequency difference between the supermodes. As shown in Fig. 3a, we see full periods of sinusoidal optical transmissions (fringe patterns) as $\Delta\phi$ (= $\phi_L - \phi_R$) varies from 0 to $2\pi$. The solid curves in Fig. 3a are the theory curve fits, and they all match well the experiments. For all cases of $L_f$, we observe clear non-reciprocal transmission, where the $\Delta\phi$ that corresponds to the maximum transmission for R→L ($\Delta\phi_{R\to L}$) is different than that of L→R ($\Delta\phi_{L\to R}$). We further show in Fig. 3b a linear relationship between $|\Delta\phi_{R\to L}-\Delta\phi_{L\to R}|$ and $L_f$. This result is expected because $\Delta\phi_{R\to L}$ and $\Delta\phi_{L\to R}$ are both proportional to the phase difference between the two supermodes which is also proportional to $L_f$. The experiments (circles) match well with the theory (solid) and the data all lie within the theoretical window (gray) that accounts for ±5 % process variation (slab-thickness and gap of the coupled-waveguides).

Our Ramsey-type interferometer quantifies the photon population in each photonic state. In Fig. 3a, the maximum extinction ratio of the non-reciprocal fringes is ~0.6 dB. From the theory curve fit, we obtain a population probability in the odd-mode of ~2.5 %. This number is lower than the designed 50 %, and is limited in the device by the maximum RF power reaching the modulators due to impedance mismatch between the device and the RF source.

The effective magnetic flux for photons ($B_{flux}$) of $2\pi$ corresponds to a magnetic field for electrons of ~0.2 Gauss. Here we consider an electron propagating through an AB interferometer with a device area as the area of our device (i.e., ($L_f$ + 2×(modulator length)) × (average gap of the coupled-waveguides + waveguide width)). Then, the equivalent effective magnetic field becomes $B = (h/e)(B_{flux}/area)$. Further reducing the device area, for example, by using optical resonators, would lead to a stronger effective magnetic field.

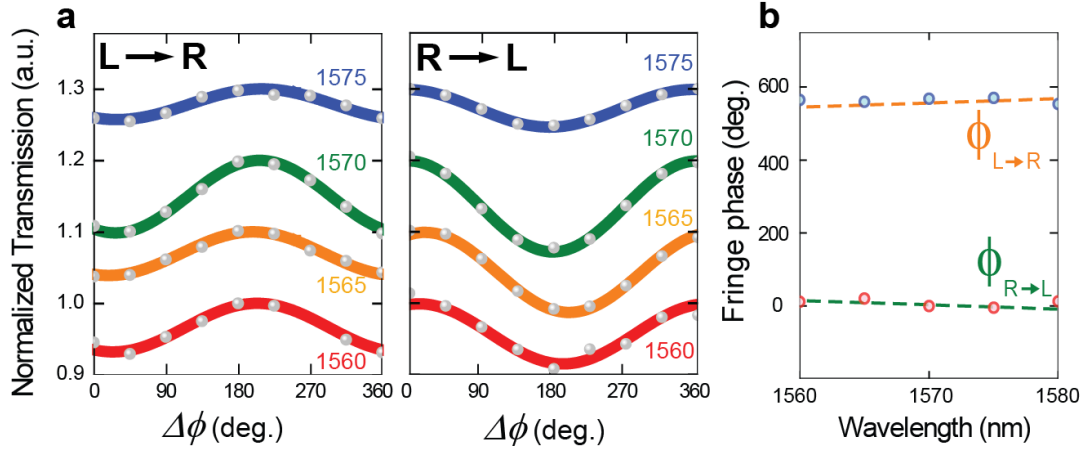

Fig. 4 **Wavelength dependence of the interference effect for the photonic Ramsey-type interferometer.** (a) Measured (gray circles) and theoretically fitted (solid) optical transmission of light traveling from left to right (L→R) and from right to left (R→L) versus Δϕ for different laser wavelengths of 1560, 1565, 1570, and 1575 nm with $L_f$ = 350 μm. (b) Theoretical (dashed line) and measured (circles) phase of the fringes for both L→R and R→L with varying wavelength.

In order to confirm that the non-reciprocal fringes result from the interference between the supermodes, we measure the fringes at different wavelengths for $L_f$=350 μm as shown in Fig. 4a. This wavelength dependence of the fringe extinction ratio is due to the dispersion of the two supermodes - the frequency difference between the supermodes perfectly matches $f_M$ at λ~1570 nm and not at other wavelengths. In Fig. 4a, we also observe a slight shift of these fringe phases as λ changes. This is expected since $\Delta n_{eff}$ is also wavelength dependent. In Fig. 4b we plot the theoretical (dashed line) and the measured (circles) wavelength dependence of $\Delta\phi_{L\to R}$ and $\Delta\phi_{R\to L}$ from 1560 nm to 1575 nm. One can see that the experiments and the theory agree well.

In summary, we use a new photonic interferometer scheme to measure a non-reciprocal phase of light and to induce an effective magnetic field for light. This scheme also quantifies the population of both photonic states. The materials and fabrication processes we use are fully CMOS compatible. The demonstration of an effective magnetic field in conjunction with recent theoretical predictions shall enable novel photonic devices including isolators and topological insulators. We also expect this work to provide stimulus to the exploration of physics and applications of effective gauge potentials and topological manipulation for photons[21-26], by demonstrating for the first time an effective gauge potential for photons that break reciprocity and time-reversal symmetry.

## Methods

**Device fabrication.** We fabricate the device using an SOI wafer. We pattern maN-2403 photoresist to mask the waveguide patter using e-beam lithography (EBL). Then, the silicon is etched down leaving a 35 nm slab. After forming the waveguide, four individual EBL steps using PMMA photoresist masks are used to implant boron (p/p+ with $1\times10^{20} / 8\times10^{17}$ cm$^{-3}$) and phosphorus (n/n+ with $1\times10^{20} / 8\times10^{17}$ cm$^{-3}$) that forms the doping profile shown in Fig. 2c. We then activate the dopants using the furnace and RTA process. After dopant activation, the waveguides are cladded with 950 nm PECVD SiO$_2$. Then we use EBL to write the mask for the vias, followed by an RIE etch through the SiO$_2$. After etching, we deposit 80 nm MoSi$_2$ and 1.6 μm of aluminum to form the electrical contact.

**Measurement and the experimental setup**. The two signals applied to each of the modulators are provided by two signal generators with the same frequency of 4 GHz to match the frequency separation of the even and odd-modes at a wavelength of 1570 nm. We choose 4 GHz as the operational frequency because both signal generators provide the maximum output at this frequency. The two signal generators, each set at 24 dBm, are synchronized to ensure a correlated phase, and for both signals, a 1 % of the RF power is dropped through a directional coupler into the oscilloscope. The correlated phases from the two signal generators are then monitored using the oscilloscope. For both signals, the other 99 % of RF power are delivered to their corresponding high speed RF probes that land on the device pads. For the optical setup, we couple light from a tunable laser into the waveguide through a lensed fiber, and the light transmission through the device is coupled out through another lensed fiber. The output light is then collected by a photodetector. Forward and backward transmissions are measured by interchanging the fiber connectors connecting the laser and the photodetector.

## Acknowledgements

This work was supported by the National Science Foundation through CIAN ERC under Grant EEC-0812072 and by the National Science Foundation under Grant No. 1202265. This work was performed in part at the Cornell Nanoscale Facility, a member of the National Nanotechnology Infrastructure Network, which is supported by the National Science Foundation. P.N. acknowledges support from Fundação de Amparo à Pesquisa do Estado de São Paulo (FAPESP grant # 2011/12140-6). We thank Dr. Gernto Pomrenke for helpful comments. We also thank MURI-Stanford for supporting this work.